\documentstyle[12pt,graphics]{article}
\title{Dynamic Localization in Quantum Wires}
\author{I.Tralle\footnote{e-mail: tralle@univ.rzeszow.pl}\\
{\em Institute of Physics, University of Rzesz\'ow}\\
{\em Al.Rejtana 16A,35-310 Rzesz\'ow, Poland}}
\date{}

\begin{document}
\maketitle
\begin{abstract}
	In the paper the dynamic localization of charged particle (electron) in a quantum wire under the external non-uniform %%@
time-dependent electric field is considered. The electrons are trapped in a deep 'dynamic' quantum wells  which are the %%@
result of specific features of the potential imposed on 2D electron gas: the scale of spatial nonuniformity is much smaller %%@
then the electron mean free path ($L_1\ll \bar{l}$) and the frequency is much greater then $\tau^{-1}$, where $\tau$ is the %%@
electron free flight time. As a result, the effect of this field on the charged particle is in a sense equivalent to the %%@
effect of a time-independent effective potential, that is a sequence of deep 'dynamic' quantum wells were the elelctrons are %%@
confined.
The possible consequeces of this effect are also discussed and  similarity with the classical Paul traps are emphasized.\\


\end{abstract}

\section {Introduction}

	To the author's knowledge, the term 'dynamic localisation' was coined by D.H.Dunlop and 
V.M.Kenkre in their well-known paper {[1]},
where they considered charged particle motion on  a 
liner chain of sites $m$ $(-\infty < m <\infty)$ under
the combined action of a time-dependent 
electric field $E(t) = E_{0}\sin\omega t$ in the direction of the lattice
and of nearest-neghbor 
coupling $V$.
The Hamiltonian considered in {[1]} is of the form:

\begin{displaymath}
H(t) = V\sum_{-\infty}^{\infty}(\mid m><m+1 \mid + \mid m+1><m\mid) - eE(t)d\sum_{-\infty}^{\infty}m\mid m><m\mid ,
\end{displaymath}
where $\mid m>$ represents  Wannier state localised on the lattice site $m$, $d$ is the lattice constatnt, $V$ is the %%@
nearest-neigbbor intersite overlap integral and $e$ is the particle charge. 
One of the main results of the paper {[1]} is that under certain conditions charged particle has to be localised in such a %%@
lattice, namely, the mean-square displasement $<m>$ is to be bounded if the ratio of the field magnitude to the field %%@
frequency is a
root of $J_{0}$, where $J_{n}, (n=0,\pm 1,\pm 2,...)$ is 
the ordinary Bessel functions in terms of which the correspoding exponentials are expanded.

On the other hand, there is very similar but somewhat different type of charged particle localisation which also could be %%@
called 'dynamic'. It is the charged particle confinement in a nonuniform rapidly oscillating 
field which for the first time was considered in the framework of classical physics by W. Paul 
and M. Raether
{[2]}. This seminal work was the starting point for the development of something which 
was called 'Paul traps' and
which later became one of the main tools for trapping a few and even 
single ions for the purposes of ions cooling and high-resolution spectroscopy.

The point is that, as is known from calssical electrodynamics, there are no absolute maxima and
minima of a potential in the electric field free of charges. As a consequence, the localisation
of charged particles in this field is impossible if one means under localisation such a state
where the particle with energy less than some definite value cannot leave the bound region under
any initial conditions (Earenshow Theorem).  However, as it was shown by W. Paul and M. Raether {[2]}, the localisation of a %%@
particle in a
\em non-uniform high-frequency \rm electromagnetic field is possible.
It turns out that under definite conditions the force acting on a particle can be represented by means of some effective %%@
time-independent potential. The force also does not depend on the sign of the particle charge and
the particle can be localised in the space region where effective potential  has its minimum.

It is quite clear that consistent description of ions trapping and their subsequent cooling in
Paul traps needs to be treated in the quantum mechanical framework. Such quantum-mechnical analysis
of the particle motion in a rapidly oscillating field has been done by R. J. Cook, D.G. Shankland 
and
A.L. Wells {[3]}. Having supposed the amplitude of the particle wave function slowly varying function of time (to compare %%@
with short time interval $2\pi/\omega$, where $\omega$ is the field frequency),
and making average over time much greater than $2\pi/\omega$, the authors of {[3]} have shown that indeed, the particle %%@
motion in such rapidly oscillating field could be considered as if it occured in an effective \em time-independent \rm %%@
potential $V_{eff}= \bigtriangledown V(x)\cdot\bigtriangledown V(x)/4m.\omega^{2}$,
where $V(x)$ is the space-dependent part of the initial potential, $m$ is the particle mass. 
This
result is equivalent to the classical one, since the force is the negative gradient of the 
potential: $\vec{f} =-\bigtriangledown V(x)$.

Somewhat later author also considered charged particle motion in a rapidly oscillating field in
the quantum mechanical framework {[4,5]}. Contrary to {[3]}, in {[4,5]} first, particle motion 
in a
\em space-periodic high frequency field \rm was considered and second, there was no any averaging 
over time.
It turns out, that under certain conditions the solution of corresponding Scr\"odinger 
equation is asymtotically exact: if the spatial preiod is small enough and field frequency is 
sufficiently high,
the effect of the field on the charged particle is in a sense equivalent to the effect of the time-independent effective %%@
potential, that is a sequence of deep 'dynamic' 
potential wells where the electrons are confined. However, in {[4,5]} only one-dimensional model was considered, which is %%@
certainly not up to what could be required, since it is not relevant in full to the real world: in a real world charged %%@
particle always has a possibility to escape to other dimensions.
Hence, if one would like to treat the results of {[4,5]} seriously, one should 
point out physically meaningful situation where such confinement could be observed. So, the aim 
of this paper is to consider more realistic model of such micro-scale dynamic localisation 
and to discuss its possible consequences.

\section{One-dimensional model}

	Since the one-dimensional model plays an important role in further considerations, we discuss it here breafly putting %%@
the details of calculations into Appendix 1.

So, let us have a charged particle (an electron, for definiteness) moving in a potential of the 
form
$V(x,t)=V(x)\cos\omega t$ where $V(x) = V(x+n\ell)=V_{0}\cos k_{0} x$, and $\ell$ is the 
spatial period,
$n$ is the integer, $k_{0} = 2\pi/\ell$, $V_{0}$ is the amplitude. Suppose the 
movement of electron is
governed by the one-dimensional Schr\"odinger equation
\begin{equation}
i\hbar \frac{\partial\psi}{\partial t} = (-\frac{\hbar^{2}}{2m}\frac{\partial^{2}}{\partial x^{2}}
+ eV(x)\cos\omega t)\psi,
\end{equation}
where $e$,$m$ are the electron charge and mass, repectively.
Suppose for a moment that only the potential-energy term were present on the right-hand side of (1).
Then the solution of this equation would be
\[\psi(x,t)= \varphi(x,0)\exp(-iV(x)\sin(\omega t)/\hbar\omega).\]
This shows that dominant effect of the potential is to add an oscillating phase factor to the 
wave
function $\psi$. Hence, as it was suggested in {[3]}, it is natural to look for a solution 
to Eq (1)
of the form
\begin{equation}
\psi(x,t) = \varphi(x,t)\exp(-iV(x)\sin(\omega t)/\hbar\omega).
\end{equation}
Substituting (2) and $V(x) = V_{0}\cos k_{0}x$  into (1) we have
\begin{equation}
i\hbar \frac{\partial\varphi}{\partial t} = [H_{0} + H_{1}(t) + H_{2}(t)]\varphi,
\end{equation}
where
\[H_{0} = -\frac{\hbar^{2}}{2m}\frac{\partial^{2}}{\partial x^{2}} + (4m\omega^{2})^{-1}(eV_{0}k_{0}x)^{2}
\sin^{2}k_{0}x,\]
\[H_{1} = -\frac{(eV_{0}k_{0})^{2}}{4m\omega^{2}} \sin^{2}k_{0}x\cos2\omega t ,\]
\[H_{2} = -\frac{i\hbar}{m\omega}(ek_{0}V_{0}\sin k_{0}x\frac{\partial}{\partial x} + \frac{1}{2}
eV_{0}k^{2}_{0} \cos k_{0}x)\sin\omega t .\]

Now it is clear that averaging over time interval much greater than $2\pi/\omega$, or in other
words, over the time interval characterising changes of the amplitude $\varphi(x,t)$ and substituting
$\sin\omega t$ and $\cos 2\omega t$ by their average values of $0$, one would have the particle 
motion
in effective potential
\[V_{eff} = \frac{(eV_{0}k_{0})^{2}\sin^{2}k_{0}x}{4m\omega_{2}}=\frac{\bigtriangledown V(x)
\bigtriangledown V(x)}{4m\omega^{2}},\]
that is, the result of {[3]}. But we do not restrict the consideration only to that rough approximation;
instead, we do carry out the subsequent analysis in two steps: first, we consider the equation 
with
time-independent right-hand side (i.e., with the Hamiltonian $H_{0}$) and then, 'turning on' 
the terms
$H_{1} + H_{2}$, look at the consequences of it. We have then the equation
\begin{equation}
i\hbar \frac{\partial\varphi}{\partial t} = (-\frac{\hbar^{2}}{2m}\frac{\partial^{2}}{\partial x^{2}} +
\frac{(eV_{0}k_{0})^{2}}{4m\omega^{2}}\sin^{2} k_{0}x)\varphi .
\end{equation}
Introducing the new variables $\tilde{x}=k_{0}x$ and $\varphi(\tilde{x},t) = \bar{\varphi}(\tilde{x})
\exp(-i\hbar^{-1}\epsilon_{n}t)$
we obtain
\begin{equation}
\frac{d^{2}\bar{\varphi}}{d\tilde{x}^{2}} + (a_{n} + 2q\cos 2\tilde{x})\bar{\phi} = 0 .
\end{equation}
Here $a_{n} = -\beta(1-\varepsilon_{n}/\alpha)$, $q = \beta/2 = m\alpha/\hbar^{2}k^{2}_{0}$ are 
dimensionless quantities; $\alpha = (eV_{0}k_{0})^{2}/8m\omega^{2}$ and $\varepsilon_{n}$ are 
the
quantities of the dimension of energy.

Equation (5) is the Mathieu equation, the parameter $a_{n}$, considered as a function of $q$, is the  eigenvalue related to %%@
the corresponding Mathieu function. The Mathieu functions are the eigenfunctions
of the Sturm-Liouville problem for the equation (5) and the boundary conditions either
\begin{displaymath}
\bar{\varphi}(0) = \bar{\varphi}(\pi)=0,  \; \rm for\; \em se_{n}(\tilde{x},q) \end{displaymath} or 
\begin{equation}
\frac{d\bar{\varphi}(0)}{d\tilde{x}} = \frac{d\bar{\varphi}(\pi)}{d\tilde{x}}=0,\;\rm for\;
\em ce_{n}(\tilde{x},q)
\end{equation}

Which of the two conditions and hence, which of the two Mathieu functions, $se_{n}(\tilde{x},q)$ 
or $ce_{n}(\tilde{x},q)$
is to be chosen, depends on the particular situation; it'll be discussed a little bit later.
Now let us make some estimates. Suppose the characteristic length (spatial period) of the potential
$\ell \sim 10^{-4} cm$, $k_{0}=2\pi\times 10^{4} cm^{-1}$, $\omega \sim 10^{10} Hz$, $V_{0} =1 V, q=(1/8)(eV_{0}/\hbar\omega)%%@
^{2}= 2.89\cdot 10^{9}$.
Hence, at so large values of $q$ one can use the asymptote of the eigenvalues [6,7]
\begin{displaymath}
a_{n}\sim -2|q| + 2(2+1)|q|^{1/2}; \; n= 0,1,...
\end{displaymath}
and for the energy values we have
\begin{equation}
\varepsilon_{n} = (2n+1)|q|^{-1/2}\alpha .
\end{equation}
Note that the obtained 'spectrum' is equidistant with the distance between the levels $\Delta
\varepsilon_{n,n-1} = e\hbar V_{0}k_{0}^{2}/4m\omega$, which for the chosen values of paprameters
is equal to $\Delta\varepsilon_{n,n-1}= 7.33\times 10^{-15} erg = 4.5\, meV $ .

Let us consider now the solution of (3) with time-dependent $H_{1}(t) + H_{2}(t)$ term. In accordance
with common quantum-mechanical rules, one should search for the solution to (3) in the form
\begin{displaymath}
\varphi(x,t)=\sum_{n} b_{n}(t)\bar{\varphi}_{n}\exp(-it\varepsilon_{n}/\hbar)
\end{displaymath}
Here $\bar{\varphi}_{n}$ are the Mathieu functions from above and the time-dependent coefficients 
$b_{n}(t)$ obey the system of equations
\begin{displaymath}
i\hbar\dot{b}_{n}(t) = \sum_{m} H_{nm}(t)b_{m}(t) ,
\end{displaymath}
where
\begin{displaymath}
H_{nm}(t) = H_{nm}^{(1)}(t) + H_{nm}^{(2)}(t) =\exp(i\omega_{nm}t) \int_{0}^{\pi(2\pi)}
\bar{\varphi}_{n}(H_{1}(\tilde{x},t) + H_{2}(\tilde{x},t))\bar{\varphi}_{m}d\tilde{x}.
\end{displaymath}
Here $\omega_{nm}=(\varepsilon_{n} - \varepsilon_{m})$.

Depending on whether the functions are $\pi$ or $2\pi$-periodic, in the last integral the upper
limit is $\pi$ or $2\pi$.

Using the orthogonality and normalization of Mathieu functions, one can prove that $H_{nm}^{(1)}
=0$ if $\bar{\varphi}_{n}$ and $\bar{\varphi}_{m}$ are of different parity and $H_{nm}^{(1)}
\neq 0$ if $\bar{\varphi}_{n}$ and $\bar{\varphi}_{m}$ are of the same parity, that is both
are odd or both are even.

Just the opposite, $H_{nm}^{(2)}\neq 0$ only if $\bar{\varphi}_{n}$ and $\bar{\varphi}_{m}$
are of different pariry and hence, one can treat the terms $H_{nm}^{(1)}$ and $H_{nm}^{(1)}$
independently. It can be shown (see Appendix 1) that the system of equations which the coefficients
$b_{n}$ are to obey, is asymtotically of the form
\begin{equation}
i\hbar\dot{b}_{n}(t) = \sum_{m} H_{nm}^{(1)}(t)b_{m}(t) = \sum_{m}M_{nm}b_{m}(t)exp(i\omega_{nm}t).
\end{equation}
Here $M_{nm}$ is the real matrix explicit form of which is given in Appendix 1.
Using another remarkable 
asymptotes of the coefficients of corresponding expansions representing
Mathieu functions 
(for details see Appendix 1) and substituting $C_{n} = b_{n}\exp(i\lambda_{n})$,
where $\lambda_{n}$ are the parameters to be determined, one has
\begin{equation}
i\hbar\dot{C}_{n}=\sum_{m}M_{nm}C_{m}\exp [it(\omega_{nm}+\lambda_{n} - \lambda_{m})] .
\end{equation}
As it is shown in Appendix 1, an appropriate choise of $\lambda_{i}$ removes the time dependence
from $H_{1}(t)$, yeilding the solution $\hat{C} = \hat{C}(0)e^{-i\hbar^{-1}tM}$, 
where $\hat{C}(0) = (C_{1}(0),
C_{2}(0),...,C_{N}(0))$ is the row matrix and $N$ is the maximum 
number of states (quantum levels)
to be included in consideration.

Since the matrix $M_{nm}$ is Hermitian, the solution of (9) finally has the form
\begin{displaymath}
\hat{C}(t) = \rm diag [\em e^{(i/\hbar)\gamma_{1}t},e^{(i/\hbar)\gamma_{2}t},...,e^{(i/\hbar)\gamma_{n}t}]\hat{C}(0),
\end{displaymath}
where diag[...] is the diagonal matrix, $\gamma_{i}$ are the eigenvalues of $M$, which are not
necessary all to be different.
So, despite the fact the Hamiltonian in (3) is explicitly 
time-dependent and the energy, generally speaking, is not conserved, the asymptotic properties of Mathieu function caused by %%@
very large value
of $q$ and the small ratio of $\omega/\omega_{n,n+1}$ cause the charged particles to be in
the states with the energies $\varepsilon_{n}$ and probabilities, corresponding to their initial distribution.

The physical explanation of this consisits of the fact that, since $\omega \ll \omega_{n,n+1}$, 
the
perturbation $H(t)$ can be regarded as adiabatic. But as it is known from quantum mechanics,
the adiabatic perturbation cannot cause the transitions between the states of the discrete spectrum
{[8]}. The adiabaticity condition is that the changes of the interaction energy during a period of oscillations in quantum %%@
systems are much smaller than the absolute values of the energy differences between the corresponding states {[8]}:
\begin{displaymath}
|\omega_{nm}^{-1}\frac{d}{dt}<n|H(t)|m>| \ll \hbar\omega_{nm}
\end{displaymath}
So, the existence of the $H(t)$-term in the Hamiltonian results only in the appearance of the new phase factor of the wave %%@
function stationary states and hence we can regard the charged particle as moving in some effective time-independent %%@
potential $U_{eff} = 2\alpha\sin^{2} k_{0}x $. 
Since $\alpha $, having the dimension of energy, is much greater than the energy of the lowest 
eigenstates
of this potential  as well as of $k_{B}T$, ($k_{B}$ is the 
Boltzmann constant) for any reasonable temperature, it is quite obvious that for the electrons 
this potential is the sequence of a deep 'dynamic' quantum wells where the electrons are confined or localised.
However, it is also possible to provide more formal proof of electron localisation 
in such potential.
The formal proof (see Appendix 2) is based on the observation that localisation also means the  \em mean \rm electron %%@
momentum in such a state should equal to zero. 

\section{More realistic physical model: quantum wire}

	Now let us proceed to the real world keeping however in mind and trying to preserve the benifits of one-dimensional %%@
model. How could we manage to do that ? Let us undo that 'knot' in a downright fachion, restricting the electron motion in %%@
other dimensions. Consider the quantum
wire (see Fig.) containing two-dimensional electron gas (2DEG) under quantum Hall-effect conditions. 
It is known that under these conditions [9] the electron free path in 2DEG can be of the order of 100 $\mu m$ and even %%@
somewhat greater. Suppose the total length $L_x$ of the quantum wire is $L_x \sim \bar{l}= 100 \mu m$ and suppose that on the %%@
surface of the wire thin insulator layer (its thickness will be estimated a little bit later) is deposited. If the quantum %%@
wire would be made of silicon, the insulator layer could be capping oxide. Suppose also that on the surface of the %%@
'sandwich' there are the metal field electrodes in a periodic sequence with the spasing between the electrodes $L_{1}\sim 5%%@
\times 10^{-4} cm$. To produce such grating is not a problem for the modern technology; for example, in {[10]} the %%@
development of a microstructure with a distance between the field electrodes of $100 nm$ was reported.

Now by means of suitably patterned gate electrodes, one could impose on a two-dimensional electron
gas an artificial periodic potential of the form $U(x,y,t) = (U_{1}(x)\cdot U_{2}(y))\cos\omega t$,
where $U_{1}(x + nL_{1}) = U_{1}(x)$. One could also suppose $U_{2}(y)$ to be weakly dependent of
its argument, that is the first and second derivatives of $U_{2}(y)$ with respect to $y  
(0 \leq
y \leq L_{y})$ to be very small. This assumption is quite clear and acceptable ; indeed, 
the potential
applied to the gates is almost uniform along the metal strip that is, within the 
interval $(0,L_{y})$
and changes only at the very edges of the strip in the vicinities of $y=0$ 
and $y=L_{y}$. So, one
can conclude that its first and second derivatives are practically equal 
to zero almost evrywhere
within the interval $(0,L_y)$. As for $U_{1}(x)$, at the moment one 
cannot say anything escept
the artificial potential imposed on 2DEG is a periodic one. But let us 
investigate more thoroughly
the role of the insulator layer.

Since the density of charges in the insulator is much smaller than in semiconductor, it is generally
supposed to be zero and hence, one can consider the potential $\Phi(x,y,z)$ within the dielectric
as governed by the Laplace equation:
\begin{equation}
\frac{\partial^{2}\Phi}{\partial x^{2}} + \frac{\partial^{2}\Phi}{\partial y^{2}} +
\frac{\partial^{2}\Phi}{\partial z^{2}} = 0 ,
\end{equation}

If, as we supposed above, the electrodes are suitably patterned and the potential $\Phi$ is weakly
dependent on $y$, one can proceed from (10) to the equation
\begin{displaymath}
\frac{\partial^{2}\Phi}{\partial x^{2}} + \frac{\partial^{2}\Phi}{\partial z^{2}}=0
\end{displaymath}
with the boundary condition $\Phi(x,0,t_{0}) = U_{1}(x)\cos\omega t_{0} = CU_{1}(x)$. 
Since $\Phi
(x,0,t_{0})$ is a periodic function with respect to $x$, $\Phi(x,z)$ could be expanded 
into a Fourier
series of functions $\Phi_{n}(x,z) = F_{n}(z)\cos(2\pi x/L_{1})$, with the functions 
$F_{n}(z)$
obeying the equation
\begin{displaymath}
\frac{d^{2}F_{n}}{dz^{2}} = (4\pi^{2}n^{2}/L^{2}_{1})F_{n},
\end{displaymath}
and hence, $F_{n} = F_{0}\exp(-z/z_{n})$, where $z_{n} = L_{1}/2\pi n$. Now it is obvious that 
the amplitude of the $n$-th harmonic decays exponentialy with the number $n$ inreasing. The natural
question is: how thick the insulator layer should be for one could neglect the higher space harmonics, restricting the %%@
consideration only to the first one $\sim \cos k_{0}x$, where $k_{0}=2\pi/L_{1}$.
If we suppose the amplitude of the second space harmonic is ten times smaller than the first one, we can calculate by means %%@
of the formulae above, that the insulator layer thickness should be about $10 nm$. This simple analysis shows quite clear %%@
that the insulator layer filters off the higher space harmonics of imposed potential and the electrons in the bulk of the %%@
semiconductor beneath the insulator are mainly affected by the lowest ones. The higher harmonics are strongly suppressed  in %%@
the insulator and can be neglected if the insulator is sufficiently thick.

The next question to answer, is this: what the equation governs the electron motion in 2DEG
under the circumstances considered above ? The answer may seem to be obvious, but in fact it requires a special care.

It is well-known that the dynamics of electron in the semicnoductor conduction band under the
\em time-independent \rm external field can be described by an equation of the form {[9]}:
\begin{displaymath}
[E_{c} +\frac{(i\hbar\bigtriangledown + {\bf A})^{2}}{2m^{*}} + U({\bf r})]\Psi({\bf r}) =
E\Psi({\bf r}),
\end{displaymath}
where $U({\bf r})$ is the potential energy due to space-charge etc., ${\bf A}$ is the vector
potential and $m^{*}$ is the effective mass. It should be noted that the wave function
$\Psi({\bf r})$ we calculate from this equation is not the true wave function but is the smoothed 
out version that does not show any rapid variations on the atomic scale (see for details [9]).

Look again at the Fig. and suppose at first the constant potential is applied to the gate electrodes. Thus we suppose the %%@
electrones are free to propagate in $x-y$ plane and are confined in $z$-direction. Usually at low temperatures with low %%@
carrier densities only the lowest subband corresponding to the confinement is occupied and the higher subbands do not play %%@
any significant role. We can then ignore the $z$-direction altogether and simply treat the semiconductor as a two dimensional %%@
system in $x-y$ plane.

By analogy with the last equation, suppose the dynamics of the electrons  of lower subband 
in 2DEG under \em time-dependent \rm external electric field applied to the gate electrodes,
is governed by the Schr\"odinger-like equation of the form:
\begin{equation}
i\hbar\frac{\partial\Psi}{\partial t} = [\frac{\hbar^{2}}{2m^{*}}\bigtriangledown^{2}_{x,y} +
U(x,y)\cos\omega t]\Psi
\end{equation}
Here in accordance with the analysis presented above, $U(x,y) = U_{1}(x)U_{2}(y)$ where $U_{1} =
V_{0}\cos k_{0}x $ , $k_{0}=2\pi/L_{1}$ and $U_{2}(y)$ is supposed to be
$$
U_2(y)=\left\{ \begin{array}{rr}
C_1,& 0<y<L_y\\
C_2,& y=0,\,y=L_y,
\end{array}
\right.
$$
where $C_{1(2)}$ are the constants $(|C_1|>|C_2|)$. Potenial $U_2(y)$ defined in this way means the jumps of it at the edges %%@
of the sample in $y$-direction, that is the potential discontinuity on the semiconductor-vacuum interface. The last one is %%@
equivalent to the Sommerfeld 'rigid box' model of the solid state {[11]}, for which boundary conditions for the electron wave %%@
funcrion $\chi(y)$ are: $\chi(0)=\chi(L_y)=0$.

Searching for the solution to (11) in the form $\Psi({\bf r})=\chi(y)\varphi(x,t)=\frac{1}
{\surd L_y}\exp(ik_y y)\varphi(x,t)$, one gets the variable separated and together with the 
boundary conditions,
the next system of eqautions holds:
\begin{eqnarray*}
-\frac{\hbar^{2}}{2m^{*}}\frac{\partial^{2}}{\partial y^{2}}\chi(y)=\varepsilon_y\chi(y)\\
i\hbar\frac{\partial\varphi}{\partial t}=[ \frac{\hbar^{2}}{2m^{*}}\frac{\partial^{2}}{\partial^{2}x}
+ U_1(x)\cos\omega t]\varphi
\end{eqnarray*}
Here \begin{displaymath}
\varepsilon_{y}=\frac{\hbar^{2}}{2m^{*}}k_y^{2},\; k_y= \frac{2\pi}{L_y}n, n=0,\pm{1},\pm{2}...
\end{displaymath}

With $U_1(x)=eV_{0} \cos k_{0}x$, the second equation becomes the same as that has been considered in previous section. It is %%@
clear that in order the approach discussed above to be valid, in particular,  for the transformation (2) can be used, the %%@
frequency applied to the gate electrodes should be greater than the inverse time which takes for the electrones to fly %%@
between two successive gates. The last one can be estimated as follows. Let it be the electron concentration in 2DEG equal %%@
$n_s =5\cdot 10^{11} cm^{-2}$ and the Fermi velocity is of the order $v_F \sim 3\times 10^{7} cm\cdot s^{-1}$.
Then the flight time $\tau=L_1/v_F\approx (5/3)\cdot 10^{-11} s$ and hence, $\tau^{-1}\approx
(3/5)\cdot 10^{11} Hz$. It means the condition $\omega\gg \tau^{-1}$ could be fulfilled if $\omega$ would be at least of $10^%%@
{12} Hz$. However, it is worthy to note that in accordance with the
formula for $\Delta\varepsilon_{n,n-1}$, increasing the distance between the gates by a factor 2,
one could reduce the friequency of the field applied to the gates by factor 4, retaining the
spacing between the levels intact. Thus, if $L_1 \sim 10^{-3} cm$, the frequency can be of $\sim
2.5\cdot 10^{11} Hz$.

Also we should discuss which of the two boundary conditions (6) for the Mathieu equation has to be chosen. Having in mind %%@
possible experimental checking up the proposed model, it seems natural to consider quantum wire connected to the current %%@
leads. Then the second of the two conditions (6) is to be more appropriate, since the first one means there is no any current %%@
at the quantum wire-lead interfaces.

To complete analysis, make again some estimates. Suppose $L_1=10^{-3} cm, \omega\approx 2.5\cdot 10^{11} Hz, m^{\ast}\approx %%@
0.1\cdot m_e (m_e \mbox{ is the free electron mass}), V_0=10 V$. Then, $q\approx 4.628\cdot10^{8}, \Delta\varepsilon_{n,n-1}%%@
\approx 4.5 meV $  and  $\Delta\varepsilon_{n,n-1}/\hbar\omega\approx 28$ . As a result, we have come to the next conclusion: %%@
if in 
a quantum wire with the periodic gate electrodes considered above and under quantum Hall effect 
conditions, the spacing is small enough and the frequency of the field applied to the gates is 
sufficiently high, the effect of the field on the charged paricles is in a sense equivalent to the
effect of a time-independent effective potential. The 'potential relief' is a sequence of deep
'dynamic' quantum wells of the depth $ 2\alpha $ and the spectrum defined by the formula (7). The 
physical explanation of this at first sight paradoxical, fact is that, since $\hbar\omega \ll
\Delta\varepsilon_{n,n-1} $, the perturbation $H(t)$ can be regarded as adiabatic and hence, it 
cannot cause the transitions between $\varepsilon_{n}$- states.

Also it is curious that, despite of the initial potential $U(x)=V_0\cos k_{0}x$ does not look like
the harmonic one of the standard quantum mechanical textbooks, the resultant specrtum turns 
out to be equidistant just like for harmonic potential $\sim x^{2}$ it is. This fact also can be 
easily explained. Indeed, at chosen values of the structure parameters, the main dimensionless
parameter $q$ which defines all the physics, is of the order of $4.628\cdot 10^{8}$ and hence, the
effective time-independent potential $U_{eff}=2\alpha\sin^{2}k_{0}x$ for small values of $x$ 
looks like the harmonic one.

\section{Discussion and conclusion}

	In the paper the dynamic localisation of charged particle (electron) in a quantum wire under the external non-uniform %%@
time-dependent field is considered. The electrons are trapped in a deep 'dynamic' quantum wells (stricktly speaking, quantum %%@
dots, since the electron 'energy spectrum'
turns  out to be completely discrete in there) which are the result of specific features of the potential imposed on 2D %%@
electron gas: the scale of spatial nonuniformity is much smaller then the electron mean free path ($L_1\ll \bar{l}$) and the %%@
frequency is much greater then $\tau^{-1}$, where $\tau$ is the electron free flight time. As a result, the effect of this %%@
field on the charged particle is in a sense equivalent to the effect of a time-independent effective potential, that is a %%@
sequence of deep 'dynamic' quantum wells were the elelctrons are confined.

Certainly the next question immediately arises. Whatever large, the dimensionless parameter $q$ is not however, \em %%@
infinitely \rm large. But mathematically it is just not we need: we should have $q\rightarrow\infty$, in
order to neglect some part of the perturbation. So, what are the consequencies of the $q$ is finite  ? 

It is quite clear that the picture sketched above is just an approximation.  Since $q$  though large, is finite, neglected %%@
part of the perturbation would cause the transitions
between the levels. However, the rate of the transitions is small compared to the inverse field frequency. There are should %%@
be about 20 or 30 of field oscillations for the transition probability reaches the unit. The complete analysis of charge %%@
carriers transport in the 'dynamic' quantum wells under such transitions is beyond the scope of the paper; we plan to discuss %%@
it in the next publication. Here we note only that the localisation considered above seems to lead on one hand, to the %%@
current suppression in a quantum wire and on the other, to the emerging of current peaks on its current-voltage %%@
characteristic (so called,  \em I-V \rm curve). These peaks are due to electron delocalisation, or escaping from the wells  %%@
by means of resonant tunnelling which seems to be quite possible, since the tunnelling time (do not confuse it with the %%@
tunnelling rate which is proportional to the inverse tunnelling probability) is much smaller then the rate of transtions %%@
between the $\varepsilon_n$-states. Indeed, omitting the subtle question of prooper definition of tunnelling time (see %%@
discussion on that subject in  {[12]}), we could estimate it roughly as $\tau_{tl}\sim s_{br}/v_{g}$, where $s_{br}$ is the %%@
width of the 'barrier' between two successive dynamic wells and $v_{g}$ - group velocity which  could  be supposed equal to %%@
Fermi velocity. Supposing $s_{br}$ to be approximately ten times smaller than $L_1$ (which seems quite reasonable), we get %%@
then $\tau_{tl}$ 200 or even 300 times smaller then the transmission rates. Hence we conclude that for tunnelling electrons %%@
the 'potential relief' they encounter on their way indeed looks like the static one. 

Putting it into plain words, one can say that since the electron life time in $\varepsilon_n$- states is long enough (the %%@
transition rates as we remember, slow down), no wonder that just for these states probability of tunnelling increses.

Suppose we have a sequence of barriers on which partciles are incident and suppose the energy of particles is smaller than %%@
the height of barriers. Then the barriers are practically impenetrable for almost all particle energies; however, for certain %%@
discrete energies and respective energy level widths particle can pass through the barriers without any reflection. This is %%@
the resonant tunnelling phenomenon [13,14]. It is quite remarkable that from the mathematical point of view the transparency %%@
of the barriers for these discrete energies does not depend on the width of the barriers if the barriers are identical [13]. 

Return now to the stationary equation (4) and rewrite it in the form:
\begin{equation}
\frac{\hbar^{2}}{2m^{*}}\psi^{\prime\prime}+(\epsilon - U_{eff}(x))\psi,
\end{equation}
where
$$
U_{eff}(x)=\left\{ \begin{array}{ll}
2\alpha\sin^{2}k_0x ,& x\in [0,L_x]\\
0,& \; x< 0,\,x>L_x.
\end{array}
\right.
$$

The tunnelling problem consists of finding all the solutions of (12) parametrically dependent on $\epsilon \in (0,\infty)$ %%@
and behaving as 
$$
\psi=\left\{ \begin{array}{ll}
\exp(ipx)+r(p)\exp(-ipx) & \;  x < 0, \\
t(p)\exp(ipx) \; & \;  x>L_x ;
\end{array}
\right.
$$
(the case of a normalized particle beam incident from the left); $p^{2}=\epsilon$; $t(p),\, r(p)$ stand for transition and %%@
refelection amplitudes, respectively.

We introduce now the outgoing propagator $G^{+}(x,x^{\prime};p)$ and define the electron wave function along the internal %%@
region $0\leq x \leq L_x $ by $\psi(p,x)=2ipG^{+}(0,x;p)$. Then the transmission amplitude is $t(p)=2ipG^{+}(0,L_x;p)\exp (-%%@
iL_x)$. Near an isolated pole $p_n = \delta_n - i\gamma_n$ it is possible to write the propagator in the form [13]:
\begin{equation}
G^{+}(x,x^{\prime};p)\approx \frac{\bar{\varphi}_n(x)\bar{\varphi}_n(x^{\prime})}{2p(p-p_n)}.
\end{equation}
Here $\bar{\varphi}_n(x)$ are the eigenfunctions of the Hamiltonian in the left-hand side of (12), $p^{2}_n =\varepsilon^%%@
{\prime}_n - i\Gamma_n/2, \varepsilon^{\prime}_n=\delta^{2}_n- \gamma^{2}_n, \Gamma_n=4\delta_n\gamma_n$. Using (13) and %%@
taking into account that $\delta_n\gg\gamma_n$, for the transmission coefficeints  $|t(p)|^{2}$  we have
\begin{displaymath}
|t(p)|^{2}=\frac{p^{2}|\bar{\varphi}_n(0)|^{2}|\bar{\varphi}_n(L_x)|^{2}}{\delta^{2}_n[(p-\delta_n)^{2}+
\gamma^{2}_n]}.
\end{displaymath}

Let us notice that in accordance with {[13]}, $\gamma_n=(|\bar{\varphi}_n(0)|^{2} + |\bar{\varphi}_n(L_x)|^{2})/2I$, where $I%%@
\approx 1$. It is also clear from the formula for  $|t(p)|^{2}$ why we choose the second of the two boundary conditions (6).

If the potential in (12) possesses only two maxima (consists of two barriers), i.e. $L_x=2\pi/k_0$, then, since the barriers %%@
are symmetric, i.e. $|\bar{\varphi}_n(0)|^{2}= |\bar{\varphi}_n(L_x)|^{2}$, we have
\begin{displaymath}
\lim_{p\rightarrow\delta_n (\epsilon\rightarrow\varepsilon^{\prime}_n)}|t(p)|=1
\end{displaymath}

Now, let the number of identical barriers be $m > 2$; then as it was proved in {[14]}, if any arbitrary pair of barriers is %%@
completely transparent at some energy $\epsilon = \varepsilon_n$, the whole structure consisted of $m > 2$ replicas is also %%@
completely transparent at this energy. 

Thus it may be said that just like in the static quantum wells, in the sequence of 'dynamic' ones, the electron resonant %%@
tunnelling also seems possible because the barrier transmition coefficient has maxima if the electron energy $\epsilon%%@
\rightarrow\varepsilon_n$.
 
To our mind, such 'dynamic' quantum wells would have even some advantages in comparison with static semiconductor %%@
heterostructures, since their 'spectrum' could be controlled by the external high-frequency electric field applied to the %%@
gate electrodes. 

In previous section we estimated the field frequecny $\omega$ as to be equal $\sim 2.5\times 10^{11} Hz$. This value is %%@
determined mainly by the electron mean free path which under quantum Hall effect conditions can be about $100 \mu m$ and %%@
hardly could be greater. To apply the field of such frequecny to the microstructure, whatever possible, is difficult problem %%@
to solve and it is interesting to know whether it is possible to reduce the frequency. We plan to consider this really %%@
intriguing problem in the next publication, here instead, we discuss another interesting question: how the effect considerd %%@
in the paper relates to the quantization of charge carrier energy spectrum in the \em uniform \rm electric field, that is to %%@
Wannier-Stark effect.

In 1960, G.H. Wannier {[15]} studied electronic states in the presence of a uniform electric field and found that eigenstates %%@
are localised along the direction of the electric field and have quantized energy levels $\epsilon_n = neEd$, where $n$ is an %%@
integer, $E$ the electric field and $d$ the lattice period along the electric field. 

The charge carriers in a crystal in the external electric field, in order to pass from the domain of the lower potential to %%@
that one of the higher potential, should gain an additional energy. Hence, the electrons in a crystal in an external electric %%@
field are to be localised in a domain with the characteristic length of the order of $2\Delta_s/eE$, where $\Delta_s$ is the %%@
subband width. As a result, the electron energy spectrum in a pure crystal whithout impurities would resemble the ladder %%@
which is called Stark, or Wannier-Stark ladder. Therefore, the electron dynamics along the field direction, say $x$-%%@
direction, can be characterised by the $k_x$-momentum only for those values of $E$, for which localisation length are much %%@
greater than $d$. This field range, $2\Delta_s\gg eEd$ is called classical. If the field increases, the situation may come %%@
into being when the localisation length becomes of the order of $d$. It is clear that the energy dependence on the qausi-%%@
momemtum $k_x$ is no longer parabolic, the electron turns out to be trapped in the potential well and its dynamics %%@
practically does not depend any further on the crystal lattice parameters. The corresponding field range, $2\Delta_s\geq eEd%%@
$ is called quantum.

The localised Stark ladder states are associated with Bloch oscillations {[16]}. The frequency of Bloch oscillations is %%@
$\omega_B = eEd/\hbar$. In order to observe the Bloch oscillations, one should have the next condition fulfilled: $\omega_B%%@
\tau_p \geq 1$, where $\tau_p$ is the momentum relaxation time. Since usually in the 3D semiconductors $\tau_p$ is of the %%@
order of $10^{-13} s$, the field should be about $2\cdot 10^{5} V\dot cm^{-1}$ in order to satisfy this condition. Usually %%@
even the lower fields cause the electric breakdown and that is why the Bloch oscillations were not observed in bulk %%@
semicnoductors.  

There have been also a number of other controversial arguments on the Stark ladder states. Some claimed {[17]} that such %%@
ladder states cannot exist because they will decay rapidly due to Zener tunnelling caused by interband mixing. Some others %%@
claimed that such interband effects are totally absent and no Zener tunnelling is observable {[18]}. Others claimed that %%@
Zener tunnelling is possible in Stark ladders {[19,20]}.

A number of experiments has been done in order to observe a Stark ladder in bulk materials, but no clear evidense was found. %%@
Only weak oscillations of conductivity observed in ZnS was attributed to the hopping motion of electrons between Stark ladder %%@
states {[21]}. The failure of observation is attributed to scattering by impurities or phonons, which prevent the %%@
accelaration of electrons up to a Brillouin-zone edge. However, in superlattices electrons can be accelerated easily to the %%@
edge of the Brillouin zone before they are scattered because of the large lattice constant. Since then, a number of %%@
experiments on Wannier-Stark effects have been performed using various techniques (see {[22]}).

It is clear that the effect discussed in the paper though different, resembles to some extent Wannier-Stark localisation , %%@
because it is the electron localisation in \em high-frequency non-uniform \rm electric field. Since the localisation length %%@
is much greater than $d$ and the momentum relaxation time in quantum wires under quantum Hall effect conditions is much %%@
greater than in bulk semiconductor, one can hope that these states could be observable just like Wannier-Stark states are %%@
observable in the superlattices.

For example, it is known that magnetophonon resonance (MPR) is observable in the superlattices {[23]}. The MPR is due to the %%@
oscillating behaviour of the electron density of states arising as a result of the Landau quatization. The magnetophonon %%@
resonance appears every time when the phonon frequency $\omega_{LO}$ is equal to the cyclotron frequency $\omega_{c}$ in a %%@
magnetic field multiplied by small integer $n$, $\omega_{LO}= n\omega_{c}, n=1,2,..$ where $\omega_{c}= eB/m^{*}$. Thus, it %%@
seems probable, that in case of the localisation effect discussed here, the current peaks on the $I-V$ characteristics of the %%@
structure also could be observed when the optical phonon frequency $\omega_{LO}$  would be equal to $\Delta\varepsilon_{nm}/%%@
\hbar$. Here $\varepsilon_n$-states would play the same role as the Landau levels do in MPR and the last phenomenon, if %%@
observed, could be called \em electrophonon \em resonance in the high-frequency non-uniform electric field. \rm

As the final remark, it is worthwhile to return again to the beginning of the paper in order to compare the effect discussed %%@
here with the Paul trapping effect and emphasize their striking similarity.

Indeed, the ion motion in radio-frequency non-uniform electric field  to the first approximation can be represented as the %%@
sum of comparatively slow motion $\bf r\em(t)$ in some effective potential and the rapid oscillations with small amplitude %%@
$\eta (t)$ near the local equilibrium $\bf r\em_{0}(t)$:
\begin{displaymath}
\bf r\em(t)= \bf r\em_{0}(t)+ \eta(\bf r\em_{0})\cos \Omega_0 t,
\end{displaymath}
where $\Omega_0$ is the applied field frequency and the amplitude of small oscillations $\eta (t)$ is determined by the trap %%@
electric field:
\begin{displaymath}
\eta(\bf r\em_{0} )= eE(\bf r\em_{0})/M\Omega^{2}_0,
\end{displaymath}
here $M$ is the ion mass and the effective potential is of the form {[24,25]}:
\begin{displaymath}
\Phi_{eff}=\frac {eA^{2}}{M\Omega^{2}_2}(\rho^{2}_0 + 4z^{2}_0).
\end{displaymath}
Parameter $A$ is defined as $A=U_0/(\rho^{2}_0 + 2z^{2}_0)$, $U_0$ is the amplitude of the field and $\rho_0, z_0$ are the %%@
cylindrical coordinates of  $\bf r\em_{0} $. In this effective potential an ion oscillates along $z$-axis with the frequency %%@
$\bar{\omega}$ and in the $x-y$ plane with the frequency $\omega_{\rho}=\bar{\omega}/2$, where 
\begin{displaymath}
\bar{\omega}=2\sqrt{2} eA/M\Omega_0 .
\end{displaymath}
More thorough analysis {[25]}  has shown that the oscillations in the effective potential $\Phi_{eff}$ are stable if the %%@
parameter $\Omega_0/\bar{\omega}$ is sufficiently large. This analysis also showed  the ion oscillation spectrum consists of %%@
the \em infinite set \rm of discrete frequencies $n\Omega_0 \pm \bar{\omega}, n=0,\pm 1,\pm 2 ...$. 

It is easily seen that our condition $\omega \gg \tau^{-1}$ corresonds to the stability parameter  introduced in [25] and the %%@
states $\varepsilon_n$ correspond to the spectrum of ion oscillations in the Paul trap.

\newpage

\section{Acknowledgements}

	The author is greatly acknowledged to the participants of the \em Physics of a Solid State \rm seminar,  Institute of %%@
Physics, University of Rzesz\'ow for the dsicussions and especially to Prof. D. Bercha, Prof. T. Paszkiewicz and Prof. E. %%@
Sheregii for their valuable comments. Author also feels himself much indebted to Mr. G. Tomaka for helping in Fig. %%@
preparation.

\section {Appendix 1}

	Here in Appendix 1 we derive the system of equations (8) to which the $b_{n}$-coefficients 
are obeyed.
Let the Mathieu functions $\bar{\varphi}^{0}_{n}$ and $\bar{\varphi}^{0}_{m}$ be of the form (see{[6]})
\begin{eqnarray*}
\bar{\varphi}_{n}=ce_{2n}(\tilde{x},-q)=(-1)^{n}\sum^{\infty}_{r=0} (-1)^{r}A^{(2n)}_{2r}
\cos 2r\tilde{x},\\
\bar{\varphi}_{m}=ce_{2n}(\tilde{x},-q)=(-1)^{n}\sum^{\infty}_{r=0} (-1)^{r}A^{(2m)}_{2r}
\cos 2r\tilde{x}.
\end{eqnarray*}
Using the orthogonality of the functions one can prove that $H^{(1)}_{nm}=\alpha f_{nm}\cos 2
\omega t$, where 
\begin{eqnarray*}
f_{nm}=\frac{1}{2}\sum^{\infty}_{r=0}\sum^{\infty}_{s=o} (-1)^{r+s}A^{(2n)}_{2r} A^{(2m)}_{2s}(\delta
_{r-1,s} + \delta_{r+1,s})\\
=-\frac{1}{2}(A_{2}^{(2n)}A_{0}^{(2m)} + A_{4}^{(2n)}A_{2}^{(2m)} + ...+
 A_{0}^{(2n)}A_{2}^{(2m)} + A_{2}^{(2n)}A_{4}^{(2m)} + ....).
\end{eqnarray*}

The evaluation of the sum in the brackets requires special care, but one can note that the asymptotic
behaviour of $A^{(2n)}_{2r}$ at $q \rightarrow \infty$ does not depend on the superscript. For
example (see [6]): $\lim_{q\rightarrow \infty} (A^{(2n)}_{2}/A^{(2n)}_{0})=-2$, 
$\lim_{q\rightarrow \infty} (A^{(2n)}_{2r}/A^{(2n)}_{0})=(-1)^{r}2$ at arbitrary $n$.

Using the recurrence formula for the function $ce_{2n}$
\begin{displaymath}
a_{2n}A^{(2n)}_{0}-qA^{(2n)}_{0}=0,
\end{displaymath}
one can demonstrate that for $q=1600$ the ratio $A_2/A_0 =-1.95$ for the function $ce_0$ and 
$-1.75$ for the function $ce_2$. At the $q=4.628\cdot 10^{8}$ we have $A_2/A_0 =-1.999907$ 
for $ce_0$ and $-1.99972$ for $ce_2$, respectively.
Hence we conclude that
\begin{equation}
f_{nm}\approx -\sum_s A^{(2n)}_{2s}A^{(2m)}_{2(s-1)}
\end{equation}
and
\begin{displaymath}
H_{nm}^{(1)} =\alpha f_{nm}\cos 2\omega t\exp(i\omega_{nm}t)=\frac{1}{2}\alpha f_{nm}(\exp i(
\omega_{nm}+2\omega)t +\exp i(\omega_{nm} -2\omega)t).
\end{displaymath}

Note that at the chosen values of parameters $2\omega \ll \omega_{nm}$, and hence
\begin{displaymath}
H^{(1)}_{nm}\approx M_{nm}\exp (i\omega_{nm}t),\; M_{nm}=\alpha f_{nm}/2 .
\end{displaymath}

Suppose now that 
\begin{displaymath}
\left.
\begin{array}{ll}
\bar{\varphi}_{n}=ce_{2n+1}(\tilde{x},-q)=(-1)^{n}\sum^{\infty}_{r=0} (-1)^{r}B^{(2n+1)}_{2r+1}
\cos (2r+1)\tilde{x},\\
\bar{\varphi}_{m}=ce_{2m}(\tilde{x},-q).
\end{array}
\right.
\end{displaymath}
Then $H^{(2)}_{nm}=\mu g_{nm}\sin \omega t$, where $\mu=-i2\hbar ek_{0}^{2}(V_{0}/m\omega)$ and
\begin{displaymath}
g_{nm}=\sum^{\infty}_{r=1}rA^{(2m)}_{2r}(B^{(2n+1)}_{2r+1} + B^{(2n+1)}_{2r+3}).
\end{displaymath}
The last series is convergent, since coefficients $A^{(2m)}_{2r}, B^{(2n+1)}_{2r+k}$ are of the
order of $r^{-2}$ at $r\rightarrow\infty$. It is noteworthy that the coefficients $B^{(2n+1)}_{2r+1}$
also do not depend on the superscript at $q\rightarrow \infty$: $\lim_{q\rightarrow\infty}
B^{(2n+1)}_{2r+1}/B^{(2n+1)}_1=(-1)^{r}$. So, $H^{(2)}_{nm}$ has the form
\begin{displaymath}
H^{(2)}_{nm}=\frac{\mu g_{nm}}{2i}(\exp i(\omega_{nm} + \omega)t - \exp i(\omega_{nm} - \omega)t)
\end{displaymath}
and for $q\rightarrow\infty$ and $\omega/\omega_{n,n+1}\ll 1$, $H^{(2)}_{nm}\rightarrow 0 \;$.  
Finally we have
\begin{equation}
i\hbar \dot{b_{n}}(t)=\sum_{m}H^{(1)}_{nm}(t)b_{m}(t)=\sum M_{nm}\exp (i\omega_{nm}t)b_{m}(t),
\end{equation}
that is, Eq (8).
Owing to the absolute convergence of the series (14), one can conclude that $f_{nm}$ does not exceed
some finite quantity. Also it is known {[6]} that $A^{(2n)}_{2s}$-coefficients possess another remarkable
asymptote, namely $A^{(2n)}_{2s} \rightarrow 0$ for $n\rightarrow \infty $ and $n > q$. Hence, we
can restrict the number of states to consider by a finite number $N$. Substituting $C_{n} =b_{n}\
exp (i\lambda_{n})$ where $\lambda_n$ are the parameters to be determined, one gets
\begin{displaymath}
i\hbar\dot{C}_n =\sum_{m} M_{nm}\exp it(\omega_{nm}+\lambda_{n}-\lambda_{m})C_m.
\end{displaymath}
Thus, any choice of $\lambda_{i}$ such that for all $n,m$
\begin{equation}
\omega_{nm} + \lambda_{n} - \lambda_{m}= 0
\end{equation}
removes the time dependence from $H_{1}(t)$, yelding the solution $\hat{C}=\hat{C}(0)\exp (-i
\hbar^{-1}Mt)$, where $\hat{C}=(C_1(0),C_2(0),...,C_N(0))$ is the row matrix consisting of coefficients at the initial time %%@
$t=0$.

It is obvious that in general (16) includes up to $(1/2)N(N-1)$ equations in the $N$ unknowns
$\lambda_i$ to be solved simultaneously, so that the consistent choice of the set $\left\{\lambda_i \right\}$ is not always %%@
possible. However, we have to recollect that our 'spectrum' $\varepsilon_n$ is equidistant and in case of $N$ states the set %%@
$\left\{\omega_{nm}\right\}$ consists of $N-1$ unequal terms only. So, if we believe any of the $\lambda_i$ is equal to some %%@
definite value (zero, for example) we obtain $N-1$ simultaneous equations in $N-1$ unknowns. Thus, as it is stated in Sec.1, %%@
we have
\begin{displaymath}
\hat{C}(t) = \rm diag \em [e^{(i/\hbar)\gamma_1t}, e^{(i/\hbar)\gamma_2t},...,e^{(i/\hbar)\gamma_nt}]\hat{C}(0),
\end{displaymath}
where $\rm diag[...]$ is the diagonal matrix and $\gamma_n$ are the eigenvalues of $M$-matrix .

\section{Appendix 2}

	Here we prove the \em mean \rm electron momentum is equal to zero in the potential $U_{eff}=2\alpha\sin^{2}k_0x$, using %%@
the the Wigner function method {[26]}.

Soppose the small subsystem described by the Hamiltonian
\begin{displaymath}
H=-\frac{\hbar^{2}}{2m^{*}}\frac{d^{2}}{dx^{2}} + 2\alpha\sin^{2}k_0x
\end{displaymath}
is in the state of thermal equilibrium with the environment characterised by the temperature $T$. Thus, the density matrix of %%@
the subsystem is of the form {[8]}:
 $$
\rho(x, x^{\prime})= Z^{-1}\sum_n \bar{\varphi}^{\ast}_n (x)\bar{\varphi}_n (x^{\prime})\exp (-\varepsilon_n/k_b T),
$$
where statistical sum 
$$
Z= \sum_n\exp(-\varepsilon_n/k_BT) = Sp[(-H/k_BT)]
$$
allows the density matrix to obey the normalization condition $Sp [\rho] =1$.

The diagonal element $\rho(x,x)=P(x)$ is a probabilty for a particle to be at the point with the coordinate $x$, while $\rho%%@
(p,p)=P(p)$ is a probablility for a particle to have the moment $p$.

In the framework of classical statistical mechanics there can be introduced probability density function with the properties:
\begin{displaymath}
P(p)=\int f(p,x)dx,\; P(x)=\int f(p,x)\frac{dp}{2\pi\hbar}.
\end{displaymath}

It turns out that in quantum mechanics the role of probability density function plays Wigner function $f_W(p,x)$ defined as %%@
{[26]}:
\begin{equation}
f_W(p,x)=\int \rho(x + \eta/2, x - \eta/2)\exp ((-i/\hbar)p\eta)d\eta.
\end{equation}
Then, for a function $h(p,x)$ which is the function only of $p$, or only of $x$, the next relation holds:
\begin{displaymath}
<h(p,x)>=\int f_W(p,x)h(p,x)\frac{dp}{2\pi\hbar}dx,
\end{displaymath}
where $<..>$ stands for mean value.
Thus, in order to calculate $<p>$, we need in accordance with (17), to have the expression for the density matrix $\rho(x,x)%%@
$. With Hamiltonian considered, the equation for density matrix reads
\begin{equation}
i\hbar\frac{\partial}{\partial t}\rho(x,x^{\prime})= (H - H^{\prime})\rho(x,x^{\prime};t).
\end{equation}
Since the Hamiltonian is time-independent, one can search for the stationary solution to this equation. It is easily seen, %%@
that if one searches for the solution in the form
$$
\rho(x,x^{\prime})= C\exp (-f(x + x^{\prime}, x - x^{\prime})),
$$
it is possible, by the proper choise of the auxiliary function $f(x + x^{\prime}, x - x^{\prime})$ to subtract the potential %%@
$U_{eff}$, sutisfying in this way the equation (18).

One can check it up that the proper choise is this:
\begin{equation}
\rho(x,x^{\prime})=C\exp [-\frac{m^{*}}{a}(\frac{1}{2\hbar k_0})^{2}\sin(k_0(x - x^{\prime})) - 2a\alpha\cos(k_0(x + x^%%@
{\prime}))], 
\end{equation}
where $a$ is the arbitrary constant and $C$ is the normalizing constant defined by the usual condition
$$
\int_0^{L_x}\rho(x,x^{\prime})dx = 1.
$$

Supposing $a=i\mu$, upon expanding the exponential of the trigonometric function in terms of Bessel functions {[27]}:
$$ 
\exp(iysin\vartheta)= \sum _{n=-\infty}^{\infty} J_n(y)e^{in\vartheta},
$$
where $y=(m^{*}/\mu)(1/2\hbar k_0)^{2}$, $\vartheta= k_0\eta $ and using (17), (19), by means of direct calculation one %%@
arrives at $<p> =0$.

\newpage

\centerline{\hbox{\includegraphics{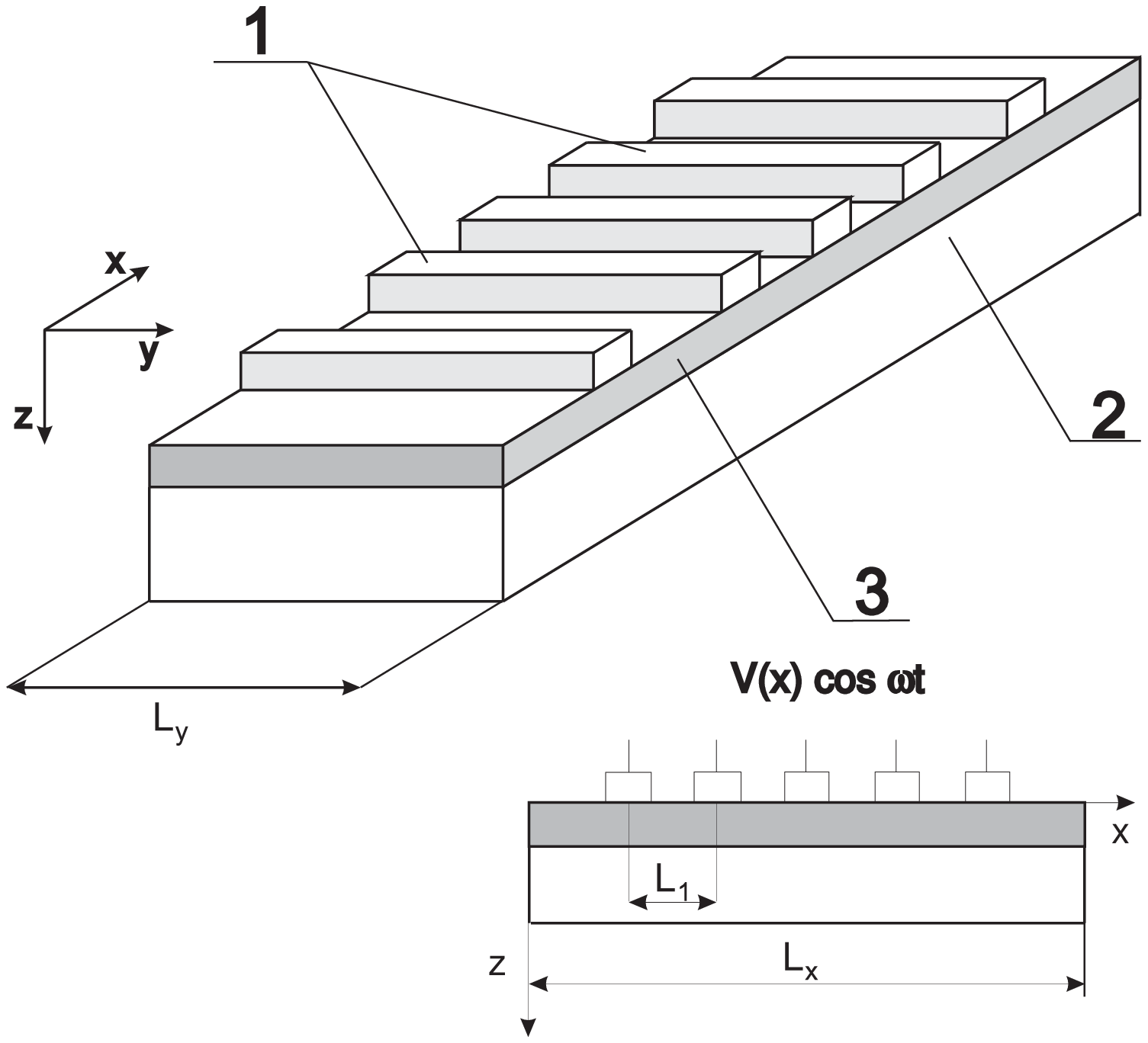}}}

\bf Figure Captions \rm

A sketch of the quantum wire with periodic gate electrodes. 1: metal gates; 2: insulator layer; 3: semiconductor  with 2DEG %%@
under quantum Hall effect conditions.
\end{document}